# Electronic structure of a wide quantum wire in a magnetic field at ν = 1

A.A. Vasilchenko

*National Research Tomsk State University, 634050 Tomsk, Russia*

The density functional theory is used to study the electronic structure of a quantum wire in a magnetic field. In a GaAs quantum wire, a critical density has been found, below which the electron density has a strong spatial inhomogeneity. The effect of the disorder potential on the electron density profile in a two-dimensional electron gas is estimated.

The results of recent experimental works [1-6] on the study of the fractional quantum Hall effect cannot be explained within the framework of existing theoretical models [7, 8]. These models also do not explain the existence of quantum Hall states at filling factors $v = 5/2$ [9], $v=3/8, 3/10$ [10, 11], $v=4/11, 4/13, 5/17, 7/11$ and other filling factors [10-12]. In addition, the results of the exact diagonalization of the many-body Hamiltonian [13] show that the wave function [7] is very far from the true ground state of two-dimensional electrons at $v = 1/3$.

In Ref. [14] we propose a new model in which the integer and fractional quantum Hall effect are explained from unified positions. We believe that droplets with a finite number of electrons can form in a two-dimensional electron gas in magnetic field at the filling factor of $v \leq 1$. In this case, the Coulomb interaction increases the energy, while the exchange interaction lowers the energy. Note that electron drops can also form in quasi-two-dimensional layers in a zero magnetic field [15–18]. Particularly noteworthy are the results of Ref. [17], which show the possibility of the formation of spin drops with the number of electrons $N = 4$ and the total spin $S = 2$.

At $v < 1$ in the one-particle approximation, some of the electronic states are unoccupied, which favors the formation of electron droplets. At $v = 1$, all states are occupied, and the electron density can be spatially homogeneous. In this work, we use the density functional theory taking into account the exchange energy in the local density approximation to study the electronic structure of a wide quantum wire in a magnetic field at a filling factor $v = 1$.

Let us consider a single quasi-one-dimensional layer of electrons in a perpendicular magnetic field. Inside the quantum wire, electrons are confined by a positively charged background with a two-dimensional density $n_p$ ($n_p$ is nonzero at $|x| \leq a/2$, where $a$ is the width of the quantum wire).

According to the density functional theory, the total energy of a multielectron system is the functional of the electron density $n(x)$:

$$E[n] = T[n] + \frac{1}{2}\int V_H(x)(n(x)-n_p)dx + E_x[n], \qquad (1)$$

where $T[n]$ is the kinetic energy of noninteracting electrons in magnetic field $B$, which is given by the vector potential $A = (0, Bx, 0)$. The second term in expression (1) is the Coulomb energy of electrons.
For the exchange energy we use the local density approximation

$$E_x = \int \varepsilon_x(n) n(x) dx, \quad (2)$$

where $\varepsilon_x = -\pi\sqrt{2\pi} L n$, $L$ is the magnetic length.

We use the atomic system of units, in which the energy is expressed in units of $Ry = e^2/(2\varepsilon a_B)$, and the length in units of $a_B = \varepsilon \hbar^2/(m_e e^2)$, where $m_e$ is the effective electron mass, $\varepsilon$ is the dielectric constant. All calculations are performed for GaAs quantum wires, for which $\varepsilon = 12.4$ and $m_e = 0.067 m_0$ ($m_0$ is the free electron mass). For GaAs we get $a_B = 9.8$ nm, $Ry = 5.9$ meV.

By minimizing functional (1) one obtains the Kohn-Sham equations:

$$-\frac{d^2\psi_k(x)}{dx^2} + \frac{(x-kL^2)^2}{L^4}\psi_k(x) + V_{eff}(x)\psi_k(x) = E_k \psi_k(x), \quad (3)$$

where $V_{eff}(x) = V_H(x) + V_x(x)$, $V_H(x) = 4\int_{-\infty}^{\infty}(n_p - n(x_1))\ln|x - x_1| dx_1$,

$$V_x(x) = \frac{d(\varepsilon_x(n)\, n)}{dn},$$

$n(x) = \int_{-k_F}^{k_F} \frac{dk}{2\pi}\psi_k^2(x)$, $k_F = \pi n_p a$.

The kinetic energy is given by

$$T = \int_{-k_F}^{k_F} \frac{dk}{2\pi}(E_k - \int V_{eff}(x)\psi_k^2(x)dx) \quad (4)$$

The nonlinear system of Kohn-Sham equations was solved numerically when $n_p\, 2\pi L^2 = 1$ (for a macroscopic system this condition corresponds to the filling factor $\nu = 1$). Figure 1 shows the electron density profiles at various values of the quantum wire width for $n_p = 7\times 10^{10}$ cm$^{-2}$. We see that the two-dimensional electron gas in wide quantum wires breaks up into stripes about 50 nm wide. At the boundary of each stripe, the density of electrons is close to zero. At small widths of the quantum wire (Fig. 1, $a = 20$ nm) the width of the electron density profile significantly exceeds the width of the quantum wire. As the width of the quantum wire increases, the electrons localize near the center of the stripe and the electron density at the boundaries of the stripe is close to zero.

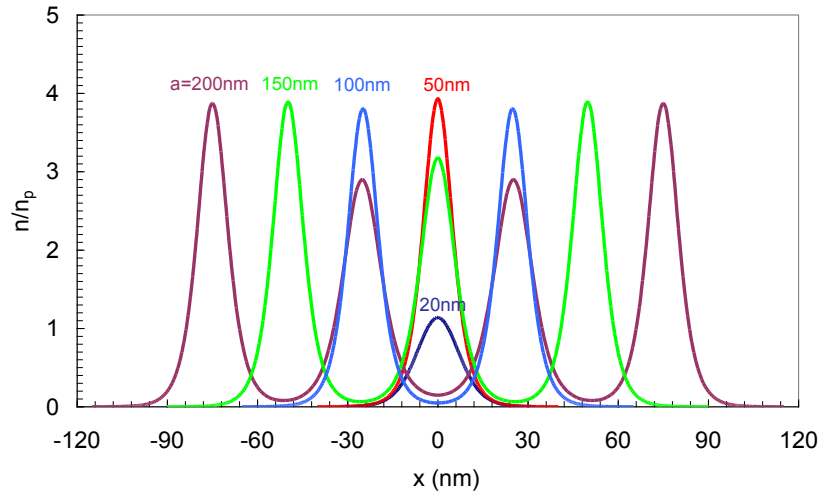

Fig.1. Electron density profiles at $n_p = 7\times 10^{10}$ cm$^{-2}$

We define the energy per electron as

$$E_1 = \frac{E}{n_p a} \quad (5)$$

For a homogeneous electron gas the energy per electron is

$$E_{1h} = 2\pi n_p - \pi\sqrt{n_p} \quad (6)$$

Figure 2 shows the energy per electron as a function of width of the quantum wire. The energy minimum is reached at $a = 60$ nm and for a macroscopic system we can assume that the two-dimensional electron gas breaks up into stripes. At $a = 60$ nm, the electron density at the stripe boundary is more than three orders of magnitude lower than in the center of the stripe (Fig. 3). This effect is associated with an increased role of exchange interaction at a low electron densities. As the width of the quantum wire increases ($a = 67$ nm, Fig. 3), the density profile broadens and the electron density at the boundary of the quantum wire is different from zero.

In the presence of a disorder potential, both isolated stripes and stripes in which electron densities overlap can be formed. Let us estimate the effect of the disorder potential on the electron density profile in a two-dimensional electron gas. For a weak disorder potential $V_d(x)$, the energy per electron is as follows:

$$E_{1d} = E_1 + \frac{\int V_d(x) n(x) dx}{\int n(x) dx}. \quad (7)$$

We assume that the two-dimensional electron gas breaks up into stripes. As an example, we consider states with $a = 60$ nm and $a = 67$ nm (Fig. 3). The energy difference between these states is about 1 meV (Fig. 2). Therefore, at disorder potential amplitudes of the order of 1 meV near the center of stripe and (or) - 1 meV near the stripe boundaries, a transition to a wider electron density profile occurs. Depending on the amplitude and scale of the disorder potential, the ground state can be either a state localized along $x$-axis or coupled regions of a two-dimensional electron gas.

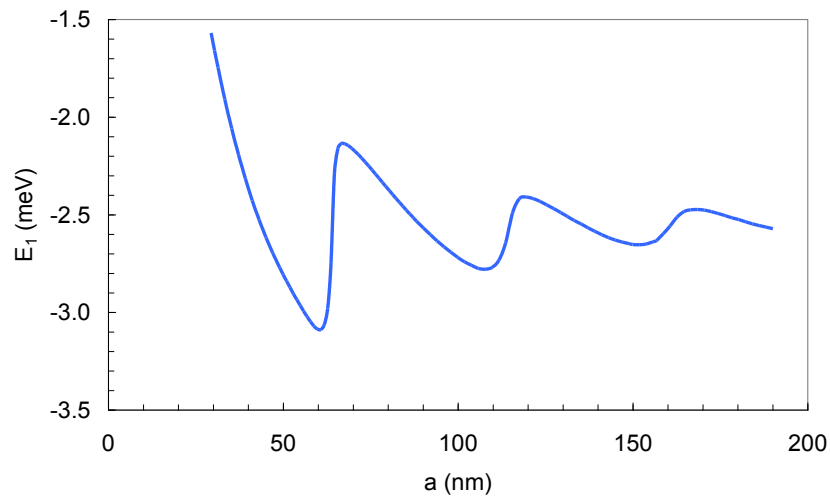

Fig.2. Dependence of energy per electron on the width of the quantum wire at $n_p = 7 \times 10^{10}$ cm$^{-2}$

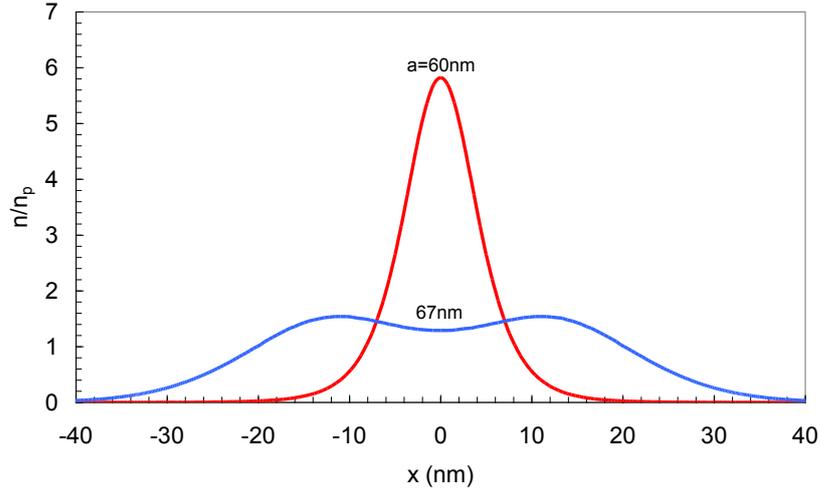

Fig.3. Electron density profiles at $n_p=7\times10^{10}$ cm$^{-2}$

As $n_p$ decreases, the first minimum in the dependence of $E_1(a)$ becomes deeper and shifts toward larger a. The calculation results for densities from $5\times10^{10}$ cm$^{-2}$ to $10^{11}$ cm$^{-2}$ show that the first minimum occurs at $n_p a \approx 4\times10^5$ cm$^{-1}$. At $n_p =10^{11}$ cm$^{-2}$, we still observe minima in the dependence of $E_1(a)$ (Fig. 4). With increasing density $n_p$, the $E_1(a)$ dependence becomes monotonic and the energy per electron decreases with increasing width of the quantum wire (Fig. 4). The energy per electron $E_1$ for the wide quantum wire is almost the same as the energy $E_{1h}$ at $n_p=10^{11}$ cm$^{-2}$ and $n_p=2\times10^{11}$ cm$^{-2}$ and lower by 0.3 meV at $n_p=7\times10^{10}$ cm$^{-2}$.

At $n_p=10^{11}$ cm$^{-2}$, the energy value of the first minimum is greater than the energy $E_{1h}$, and at $n_p=7\times10^{10}$ cm$^{-2}$, it is less than $E_{1h}$ energy. The transition to a spatially inhomogeneous liquid occurs when the energy value at the minimum $E_1 \approx E_{1h}$ and this transition occurs at the critical density $n_c \approx 8\times10^{11}$ cm$^{-2}$. For the exchange energy, we use the local density approximation, which can lead to an overestimation of the critical density value. Note that the use of a similar model for two-dimensional quantum dots [14, 19] leads to a qualitative agreement with the results of the exact diagonalization of the many-body Hamiltonian [20-25].

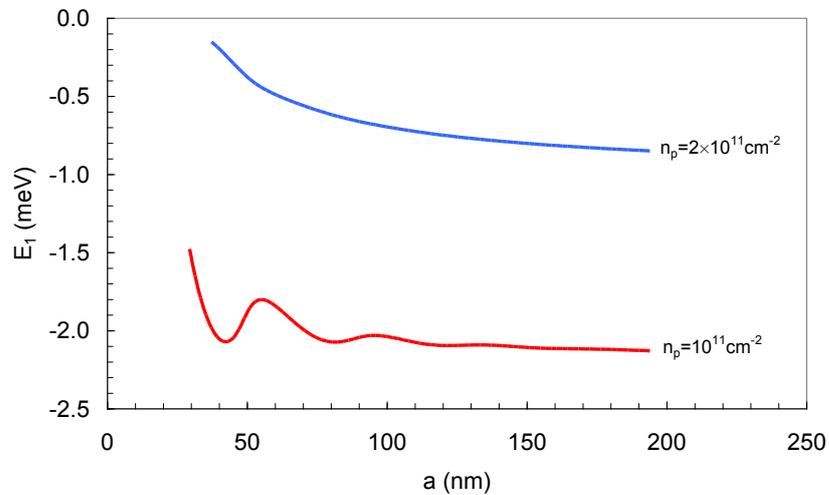

Fig. 4. Dependence of energy per electron on the quantum wire width at $n_p=10^{11}$ cm$^{-2}$ and $n_p=2\times10^{11}$ cm$^{-2}$

Figures 5 and 6 show the electron density profiles. We see that as np increases, the electron density in wide quantum wires is close to homogeneous. The energy per electron changes weakly with increasing width of the quantum wire, so a disorder potential with a scale of several $a_B$ can lead to the formation of inhomogeneous electron stripes.

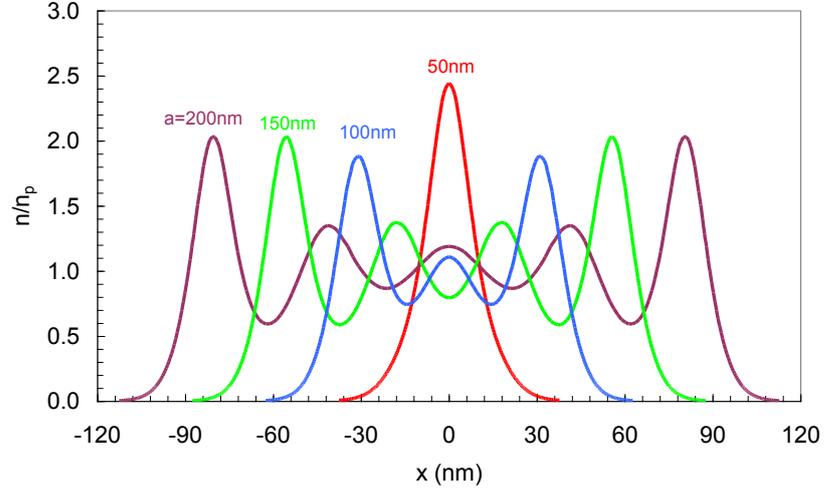

Fig.5. Electron density profiles at $n_p = 10^{11}$ cm$^{-2}$

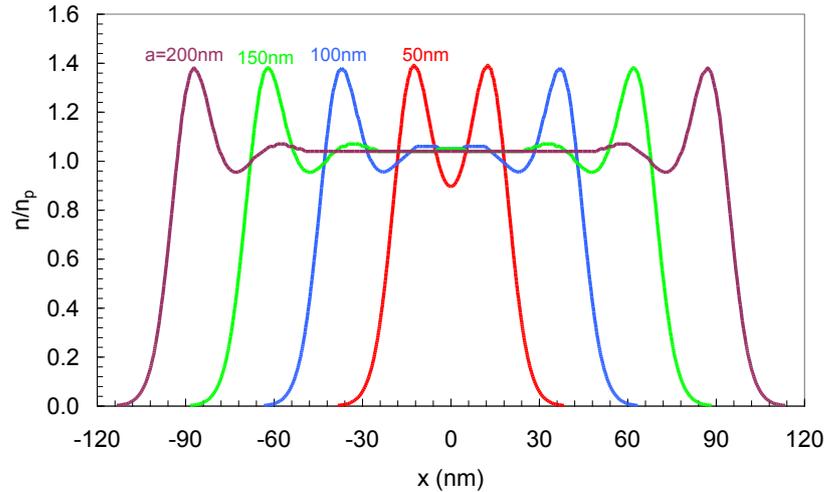

Fig.6. Electron density profiles at $n_p = 2 \times 10^{11}$ cm$^{-2}$

The symmetry chosen by us can be broken and a spatially inhomogeneous distribution of the electron density along the stripe becomes possible. In this case, electron drops with a finite number of electrons can be formed. The transverse dimensions of the drop are comparable to the width of the quantum well, therefore, to confirm this hypothesis, it is necessary to carry out calculations for a three-dimensional anisotropic drop. In the case of the formation of electron drops, the width of the Hall plateau will decrease as the number of electrons in the drop increases [26]. In a pure two-dimensional electron gas at high densities, the electron density will be spatially homogeneous and the quantum Hall effect will not be observed. However, for a more detailed study of the quantum Hall effect at

$v = 1$, it is necessary to carry out calculations taking into account electrons at the second spin level.

In conclusion, the system of Kohn-Sham equations for two-dimensional electrons in a wide quantum wire in a magnetic field with a filling factor $v = 1$ is solved numerically. It was found that in GaAs quantum wire at densities $n < 8 \times 10^{10}$ см$^{-2}$, the electron density has a strong inhomogeneity and the two-dimensional electron gas consists of uncoupled stripes. Electron density will be spatially homogeneous in a pure two-dimensional electron gas at high densities. A disorder potential with an amplitude of 1 meV or less and a scale of the order of several $a_B$ can lead to the formation of both uncoupled and coupled regions of a two-dimensional electron gas. Note that experimental studies have shown the possibility of the formation of a phase-inhomogeneous state in two-dimensional systems in a zero magnetic field [15] and in a strong magnetic field [27].

This work was supported by the State Assignment of the Ministry of Science and Higher Education of the Russian Federation (project No. FSWM-2020-0048).

**References**

[1] H. Fu, Y. Wu, R. Zhang, J. Sun, P. Shan, P. Wang, Z. Zhu, L. N. Pfeier, K.W. West, H. Liu, X. C. Xie, and Xi Lin, Nat. Commun. **10**, 4351 (2019).

[2] J. Falson, D. Maryenko, B. Friess, D. Zhang, Y. Kozuka, A. Tsukazaki, J.H. Smet, and M. Kawasaki, Nat. Phys. **11**, 347 (2015).

[3] A.S. Zhuravlev, L.V. Kulik, V.A. Kuznetsov, M.A. Khit'ko and I.V. Kukushkin, JETP Letters **108**, 419 (2018).

[4] L.V. Kulik, V.A. Kuznetsov, A.S. Zhuravlev, V. Umansky, and I.V. Kukushkin, Phys. Rev. Research **2**, 033123 (2020).

[5] L.V. Kulik, A.S. Zhuravlev, E.I. Belozerov, V.A. Kuznetsov, and I.V. Kukushkin,, JETP Letters **112**, 485 (2020).

[6] Chengyu Wang, A. Gupta, S. K. Singh, Y. J. Chung, L.N. Pfeiffer, K.W. West, K.W. Baldwin, R. Winkler, and M. Shayegan, Phys. Rev. Lett. 129, 156801 (2022).

[7] R.B. Laughlin, Phys. Rev. Lett. **50**, 1395 (1983).

[8] J.K. Jain, Composite-fermion approach for the fractional quantum Hall effect. Phys. Rev. Lett. **63**, 199 (1989).

[9] R. Willett, J.P. Eisenstein, H.L. Stormer, D.C. Tsui, A.C. Gossard, and J.H. English, Phys. Rev. Lett. **59**, 1776 (1987).

[10] W. Pan, H. L. Stormer, D. C. Tsui, L.N. Pfeiffer, K.W. Baldwin, and K.W. West, Phys. Rev. Lett. **90**, 016801 (2003).

[11] W. Pan, J.S. Xia, H.L. Stormer, *et al.* Phys Rev B **77,** 075307 (2008).

[12] V.T. Dolgopolov, M.Yu. Melnikov, A.A. Shashkin, S.-H. Huang, C.W. Liu, and S.V. Kravchenko, JETP Letters **107**, 794 (2018).

[13] S.A. Mikhailov, arXiv:2206.05152 (2022).

[14] A.A. Vasilchenko, arXiv:2209.05601 (2022).

[15] V.M. Pudalov and M.E. Gershenson, JETP Lett. **111**, 225 (2020).

[16] L.A. Tracy, J.P. Eisenstein, M.P. Lilly, L.N. Pfeiffer, and K.W. West, Solid State Commun. **137**, 150 (2006).

[17] N. Teneh, A.Yu. Kuntsevich, V.M. Pudalov, and M. Reznikov, Phys. Rev. Lett. **109**, 226403 (2012).

[18] L.A. Morgun, A.Yu. Kuntsevich, and V.M. Pudalov, Phys. Rev. B **93**, 235145 (2016).

[19] A.A. Vasilchenko, arXiv:2209.04724 (2022).

[20] P.A. Maksym and T. Chakraborty, Phys. Rev. Lett.. **65**, 108 (1990).

[21] P.A. Maksym and T. Chakraborty, Phys. Rev. B **45**, 1947 (1992).

[22] P.A. Maksym, Physica B **184**, 385 (1993).

[23] C. Yannouleas and U. Landman, Phys. Rev. B **70**, 235319 (2004).

[24] C. Yannouleas and U. Landman, Phys. Rev. B **66**, 115315 (2002).

[25] C. Yannouleas and U. Landman, Phys. Rev. B **68**, 035326 (2003).

[26] S.M. Reimann and M. Manninen, Rev. Mod. Phys. **74**, 1283 (2002).

[27] D. Maryenko, A. McCollam, J. Falson, Y. Kozuka, J. Bruin, U. Zeitler, and M. Kawasaki, Nat. Commun. **9**, 4356 (2018).